\documentclass[runningheads]{llncs}
\usepackage{graphicx}
\usepackage{cite}

\usepackage{longtable}
\usepackage{blindtext}
\usepackage{xcolor}

\usepackage{comment}
\usepackage{multirow}
\usepackage{makecell}
\usepackage[pdftex]{hyperref}

\usepackage[normalem]{ulem}
\usepackage{amsmath}
\newcommand{\stkout}[2]{\ifmmode\text{\sout{\ensuremath{#1}}}\else\sout{#1}\fi}

\begin{document}

\title{
Vulnerabilities and Open Issues of Smart Contracts: A Systematic Mapping
}
%
%
\authorrunning{G. de S. Matsumura \emph{et al.}}

\author{
    Gabriel de Sousa Matsumura\inst{1}\orcidID{0000-0002-0627-9378}
    \and Luciana Brasil Rebelo dos Santos\inst{1}\orcidID{0000-0002-5193-6218}
    \and Arlindo Flavio da Conceição\inst{2}\orcidID{0000-0002-7123-3990}
    \and Nandamudi Lankalapalli Vijaykumar\inst{2,3}\orcidID{0000-0002-9025-0841}
}
%
%
\institute{
Federal Institute of Education, Science and Technology of São Paulo, Brazil\\
    \email{sousa.matsumura@gmail.com, lurebelo@ifsp.edu.br}\\
    \and Federal University of São Paulo, Brazil\\
    \email{\{arlindo.conceicao,vijaykumar\}@unifesp.br}\\
\and National Institute for Space Research, Brazil\\
}
\maketitle              
\begin{abstract}
Smart Contracts (SCs) are programs stored in a Blockchain to ensure agreements between two or more parties. The combination of factors like the newness of the field, with less than a decade of practical use, and a high and growing popularity in industry, leads to an increase in the possibility of severe security issues occurrence. Even though, systems based on Blockchain and SCs technologies inherit several benefits, like a tamper-proof decentralized ledger and anonymous transactions. This way, even with security issues, the trend is that its popularity will increase. Due the unchangeable essence of Blockchain, failures or errors in SCs become perpetual once published, being critical to mitigate them, since the involvement of huge economic assets are common. To handle this, the academic and industrial communities have been expanding their efforts in this field, with a growth in research publications similar to exponential. Aiming at reasoning about the current state-of-art in vulnerabilities and open issues over Blockchain SCs, we conducted a systematic literature mapping over 32 primary and 18 secondary selected articles. As contribution, this work discusses and relates the selected papers, identifying gaps that may lead to research topics for future work.




\end{abstract}
\section{Introduction}\label{intro}

Software reliability is highly relevant given the dependence that modern society has on software systems. Defects can result in minor annoyances, such as not accessing social networks for a while. They may become serious, leading to severe financial losses.
Amid advances in software technology, the idea of digital financial transactions emerges, to facilitate money transactions between people anywhere in the world, without the need for mediators, such as banks or brokers. 

Blockchain \cite{schueffel2019crypto} is a technology to make these financial operations viable. It is a distributed ledger that records all transactions of a digital currency, so it is constantly updated.
In 2015, the introduction of Smart Contracts (SCs) extended the functionality of Blockchain. 

An SC is a program designed to guarantee the execution of a transaction between two or more parties. There are fundamental differences between SC and standard software, and these differences introduce new vulnerabilities and concerns. SCs are immutable computer programs stored in a Blockchain and verified and executed by some of its nodes in a distributed manner [@13] \footnote[1]{In this text, for a question of space and simplicity, we differentiate among \textit{0)} regular references, \textit{1)} primary studies (denoted with \#) and \textit{2)} secondary studies (written with @). Regular references are available at the end of the text. In order to have access to the aforementioned primary and secondary studies, the reader must check the list of papers in \textbf{\url{https://bit.ly/3tNkIG6}}~\cite{ourresultinpapers}.}.

Notice that the code of an SC cannot be modified to correct defects \cite{zhang2020framework}, i.e., once implanted, it cannot be updated. Thus, the correctness and security of SCs are necessary, as failures can cause millionaire losses, as already extensively reported in the literature \cite{zhang2020framework}.

There is an extensive number of cases where bugs and breaches in SCs led to severe economic losses. In 2016, an attack stole approximately \$50 million worth of Ether by exploiting a flaw in the Distributed Autonomous Organization (DAO) source code~\cite{gelvez2016explaining}[@13]. In 2017, the multi-signature wallet Parity had embezzlement of about \$30 million as a consequence of an attack on Ethereum. In 2018, about \$900 million was stolen due to the BEC (BeautyChain) token attack [@13]. Also in 2018, researchers performed a security analysis of nearly a million SCs using the MAIAN tool, resulting in 34,200 SCs flagged as vulnerable\cite{nikolic2018finding}. Then, with a random sample of 3,759 contracts obtained from the vulnerable contracts set, they found that 3,686 SCs had an 89\% probability of vulnerability~\cite{nikolic2018finding}[@5].

Although recent studies have proposed a huge number of tools and techniques to detect bugs in SCs, the literature shows that the development of SCs needs to be improved in many ways, in order to minimize the problems detected so far. A possible reason for this situation is the lack of a comprehensive mapping of all the existing bugs [7] and how to deal with each of them. The present paper carried out a systematic literature mapping to characterize the techniques, methods, and tools that deal with SCs' vulnerabilities and showing recent advances in this area. Thus, the results can identify eventual gaps and use this information to mitigate problems in future research.

While conducting the study, in the stage of Data Extraction and Synthesis, we have noticed that a relevant number of secondary studies were returned. So, to make a more comprehensive contribution, we have decided to include them as part of the mapping, presenting also a tertiary study. In this way, the answers to the research questions consider both primary and secondary studies.

This paper is organized as follows. Section \ref{background} shows the background to this paper. The secondary studies related to this mapping are commented in Section \ref{relatedwork}. Section \ref{researchmethod} describes the research questions and research methods adopted. Section \ref{results-and-discussions} presents the results obtained, making considerations and discussions about the outcomes. Section \ref{threats} discusses the potential threats to the research validity. Finally, Section \ref{conclusion} presents final remarks about the mapping conducted.



\section{Background}\label{background}
This section address needed concepts for understanding this work.

\subsection{Blockchain}\label{subsecBlockchain}

A Blockchain platform is a decentralized ledger. It stores transactions permanently, in a secure and auditable way~\cite{arganaraz2020detection}. Blockchain technology was popularized by Bitcoin in 2008~\cite{nakamoto2019bitcoin}, enabling reliable digital financial transactions without a trusted third party between anonymous entities.
Bitcoin merged the qualities of cryptography techniques and peer-to-peer networks. In this network, miners verify transactions as they occur, checking signatures and balances. Then, a distributed consensus algorithm is used to create a block with the validated transactions. The block is broadcasted to the entire network so that all nodes can,  after validating its correctness, add the new blocks to their copy of the ledger~\cite{alharby2018Blockchain}. The consensus algorithm maintains a trusted network without the need to trust any node in particular [@5].

\subsection{Smart Contracts}\label{subsecsmartcontracts}

A Smart Contract (SC) is software that runs on the Blockchain and represents an agreement between non-trusting participants~\cite{alharby2018Blockchain}.
The most popular Blockchain platform that supports SCs is Ethereum~\cite{buterin2014ethereum, wood2014ethereum}. However, the development of complex SCs is not trivial~\cite{alharby2018Blockchain}.
For example, the Ethereum platform uses a network of Turing complete virtual machines to validate transactions~\cite{alharby2018Blockchain}. In Ethereum, the programmer uses a high-level programming language (e.g., Solidity) to write the SCs.  The interaction between users and SCs occurs by sending a transaction to the contract address. An SC can call other SCs during its execution and can pass data as parameters, making it possible to execute untrusted code~\cite{alt2018smt}. So there is a need to ensure the correctness of the executable codes, but that is a challenging task [@13].

Before the Blockchain technology arising, although the concept of SCs already existed~\cite{szabo1997formalizing}, they were not well developed [@5]. The participation of central authorities or resource managers was a requirement at the time, limiting strongly the usefulness of smart contracts since these third parties are able to handle agreements themselves [@10, @13]. Since there is no need for a trusted third party in the Blockchain, smart contracts can be advantageously employed.

\section{Related Literature}\label{relatedwork}

In this section, we provide a concise overview of the main secondary studies related to our systematic mapping.
Regarding publication sources, Table \ref{tab:matrix-sources-sec-type} shows that journals are composed of the following study types: systematic reviews, surveys (37.5\% each), and systematic mappings (25\%). Regarding the conference publications, there was a multivocal review, a systematic review (11.11\% each) and surveys (77.78\%). Also, we have an e-print survey publication. So, in journals, systematic approaches are more often, while surveys are more often in conferences. Even though the e-print survey is not yet published in a journal or conference, it is worth mentioning that it was included since it has rich information for our paper.

For the classification of the secondary papers, we not only consider the types defined by the authors, but also the definition of each type. Systematic approaches have a precise methodology, like the one we present in the next section to minimize subjectivity and maximize reproducibility. Systematic mappings and reviews utilize a similar methodology, the main difference is the scope. Mappings are more likely to be high-level, having a  broader view about a topic and a review is more likely to be low-level, aiming at a specific perspective of a subject, often comparing two specific topics. The multivocal literature review can also have a systematic methodology, but it includes in its scope gray literature, for instance, technical reports, thesis, dissertations, etc. Finally, a survey does not utilize a precise methodology to select their primary papers. They have the tendency to have more subjectivity, increasing the odds that relevant articles may have been left behind. 
\begin{longtable}[c]{l|l|l|l|l}
\caption{Related secondary papers types and publication sources}
\label{tab:matrix-sources-sec-type}\\
\hline
\textbf{Sources and Secondary Type}    & \multicolumn{1}{c|}{\textbf{Journal}} & \multicolumn{1}{c|}{\textbf{Conference}} & \multicolumn{1}{c|}{\textbf{E-Print}} & \multicolumn{1}{c}{\textbf{Total}} \\ \hline
\endfirsthead
\endhead
Multivocal Literature Review  & 0 & 1 & 0 & 1 \\ \hline
Survey                        & 3 & 7 & 1 & 11 \\ \hline
Systematic Literature Mapping & 2 & 0 & 0 & 2 \\ \hline
Systematic Literature Review  & 3 & 1 & 0 & 4 \\ \hline
Total                         & 8 & 9 & 1 & 18\\ \hline
\end{longtable}

\begin{center}
\begin{longtable}[c]{l|l|c|c}
\caption{Focus areas identified in the related secondary papers}
\label{Table:secondary-areas}\\
\hline
\multicolumn{2}{c|}{\textbf{Areas}}
& \textbf{@Secondary Articles} & \textbf{Total (\%)} \\ \hline
\endfirsthead
\endhead
    \multirow{3}{*}{\textbf{Issues}}
        & \multicolumn{1}{c|}{\textbf{Vulnerability}} & \multicolumn{1}{c|} {1, 3, 4, 6, 7, 11, 14, 15, 17} & \multicolumn{1}{c}{38.89} \\\cline{2-4}
        & \multicolumn{1}{c|}{\textbf{Attacks}} & \multicolumn{1}{c|} {1, 7, 12, 13, 15, 17} & \multicolumn{1}{c}{33.33} \\\cline{2-4}
        & \multicolumn{1}{c|}{\textbf{External Data}} & \multicolumn{1}{c|} {8, 9, 17} & \multicolumn{1}{c}{16.67} \\\hline
    \multirow{7}{*}{\textbf{Solutions}}
        & \multicolumn{1}{c|}{\textbf{Design}} & \multicolumn{1}{c|} {4, 6, 9, 17, 18} & \multicolumn{1}{c}{27.78} \\\cline{2-4}
        & \multicolumn{1}{c|}{\textbf{Implementation}} & \multicolumn{1}{c|} {2, 4, 5, 6, 7, 8, 12, 17, 18} & \multicolumn{1}{c}{50.00} \\\cline{2-4}
        & \multicolumn{1}{c|}{\textbf{Software Testing}} & \multicolumn{1}{c|} {3, 5, 13, 14, 16, 17, 18} & \multicolumn{1}{c}{38.89} \\\cline{2-4} & \multicolumn{1}{c|}{\textbf{Formal Methods}} & \multicolumn{1}{c|} {4, 5, 6, 9, 10, 13, 14, 16, 17, 18} & \multicolumn{1}{c}{55.56} \\\cline{2-4} & \multicolumn{1}{c|}{\textbf{Machine Learning}} & \multicolumn{1}{c|} {11, 12, 17, 18} & \multicolumn{1}{c}{22.22} \\\cline{2-4}
        & \multicolumn{1}{c|}{\textbf{Tools}} & \multicolumn{1}{c|} {3, 4, 5, 6, 9, 10, 13, 14, 16} & \multicolumn{1}{c}{50.00} \\\cline{2-4}
        & \multicolumn{1}{c|}{\textbf{Monitoring}} & \multicolumn{1}{c|} {4, 9, 12, 17, 18} & \multicolumn{1}{c}{27.78} \\\hline
    \multicolumn{2}{c|}{\textbf{Blockchain Platform}}
     & 11, 12, 14, 16
     & 22.22 \\ \hline
\end{longtable}
\end{center}

In Table \ref{Table:secondary-areas}, we provide a simple categorization of the related secondary articles concerning the kind of their contributions, being common the availability of some kind of classifications and, sometimes, a taxonomy. In these issues, among papers in vulnerability scope, all of them, except [@17], mention vulnerabilities related to programming languages. Also all, but [@15] and [@17], mention vulnerabilities related to Blockchain platform and virtual machine. Considering the papers on vulnerability, we can highlight that [@15] provides a great systematization of Ethereum vulnerabilities based on the Common Weakness Enumeration that may probably be extended to other platforms. About the external data scope, all of them mention oracles, and [@17] mentions verifiable third-party and cryptography technology as possible solutions for off-chain interactions.

For solutions concerning design, we selected papers that handle modeling or specifications. For the case of implementation, solutions like design patterns [@2, @4, @8, @17], templates [@4, @17], standards [@5], more secure domain-specific languages [@5, @9, @17] and code generation [@12] are mentioned. In relation to software testing, all of the selected papers mention fuzzing test, being that [@16] also mentions test case generation. Considering the papers that mention machine learning we highlight [@12], that, unlike others, focuses on Blockchain and machine learning and, therefore, handles them more deeply, with less emphasis on SCs. With respect to monitoring SCs after deployment, papers mention runtime monitoring [@4], bug bounty [@4, @9, @17], and monitoring strategies [@18], being also worthy to mention that monitoring is often related to the update of SCs [@12, @17]. 


\section{Research Method}\label{researchmethod}

The research method adopted in this Systematic Literature Mapping (SLM) is defined based on the guidelines~\cite{keele2007guidelines}, involving three main phases: (i) Planning: referring to the pre-review activities, aiming to define the research questions, inclusion and exclusion criteria, study sources, \stkout{\textcolor{red}{re}} ssearch string, and mapping procedures to establish a protocol review; (ii) Realization: search and selection of studies, aiming at extracting and synthesizing their data; (iii) Report: final phase whose objective is to write the results and disseminate them to potentially interested parties, using the results to answer the research questions.


The research questions are listed in next section. The study selection is explained in Section \ref{subsecstudyselect}, and data extraction is presented in Section \ref{subsecdatasynthesis}.


\subsection{Research Questions}\label{subsecrqs} 


This mapping presents an overview of the current state of research on the problems/issues directly involved with SCs in the Blockchain context. Table \ref{Table:RQ}  shows the Research Questions (RQ) and the rationales considering them in this SLM.

\begin{longtable}[c]{p{1cm}|p{4cm}|p{6.5cm}}
    \caption[Research questions and their rationales]{Research questions and their rationales}  \label{Table:RQ} \\
    \hline
  \textbf{N$^{o}$} & \textbf{Research question} & \textbf{Rationale}  \\
    \hline
  \endfirsthead 
   
 
RQ1 &  When and where have the studies on vulnerabilities and open issues on SCs been published?
 &
This question is to understand whether there are specific
publication sources for these studies, and when they have been published. \\ 
\hline  
RQ2 &  What are the problems/issues with respect to SCs?
 &
 This question identifies what are the problems/issues directly related to the use of SCs in the Blockchain context. \\ 
\hline
RQ3 &  How are the problems/issues categorized or classified? & This question investigates how the problems/issues related to the use and development of SCs have been categorized/classified, checking if there is a proper taxonomy.  \\ 
\hline
RQ4 &  What are the proposed solutions to the problems/issues identified within the context of SCs? & The aim is to determine what are the proposed solutions to the problems/issues identified in RQ2.  \\ 
\hline
RQ5 &  Are there methods and tools available for the identified solutions? &  This question verifies, among the methods and tools identified in RQ4, what are the limits of using each one in terms of practical purposes.  \\ 
\hline
\end{longtable}

\subsection{Study Selection}\label{subsecstudyselect}

\paragraph{\textbf{Source and Search String.}} The search was performed in the Scopus~\footnote[2]{http://www.scopus.com} repository. We chose this database because, in the initial searches, Scopus returned a larger set of results compared to others. It was also noted that many papers found on other platforms were also returned in Scopus, considering the same search string. Finally, we defined the following search string for this mapping:

\begin{center}
    TITLE-ABS-KEY(``smart contract" w/12 (problem OR issue)) 
\end{center}

The main focus of the search was to identify papers that addressed problems/issues about smart contracts used in the Blockchain context. During the test of search strings, it was clear that the results of the string \emph {(``smart contract" AND (problem OR issue))} contemplated the papers being looked for. With the goal of narrow the resulting papers to our scope, we used the proximity operator available in Scopus, expressed by \emph {w/x}, where \emph {x} is the maximum number acceptable of words between two terms, regardless of the order of the terms. Using \emph {x=12} it was possible find previously defined control papers.


\paragraph{\textbf{Inclusion and Exclusion Criteria.}} The selection criteria are organized into two inclusion criteria (IC) and eight exclusion criteria (EC). The inclusion criteria are: (IC1) Study must include the use of smart contracts in Blockchain context; and (IC2) Study must include problems or solutions related to the use of smart contracts. The exclusion criteria are: (EC1) Study has no abstract;
(EC2) Study is just an abstract or extended abstract without full text; (EC3) Study is not a primary study, such as editorials, summaries of keynotes, and tutorials; (EC4) Study is not written in English; (EC5) Study is a copy or an older version of another publication already considered; (EC6) No access to the full paper; (EC7) Study with a publication date prior to 2015; and (EC8) Study should contain problems related to smart contracts itself, problems about smart contract applications aren't enough. 

\paragraph{\textbf{Data storage.}} We used a spreadsheet to gather all the relevant data from the returned studies in the searching phase (e.g., id, title and bibliographic reference), cataloging and storing each publication appropriately.

\paragraph{\textbf{Assessment.}}Before conducting the mapping, we checked our protocol. We defined a pre-selected set of papers, which should be present in the returned set of studies. Regarding the review process, stage 1 was conducted by one of the authors, and the papers not excluded were validated by the other authors. Stage 2 was conducted by all authors, and we equally divided the remaining papers. Throughout the review process, in cases of doubt, the papers were not excluded, and, in a meeting, the authors reached a consensus on whether or not to include the studies.


\subsection{Data Extraction and Synthesis}\label{subsecdatasynthesis}

\begin{figure}[!ht]
	\centering
		\includegraphics [width=1\textwidth]{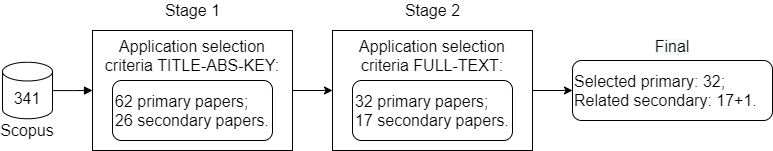}
		\caption{Search and selection SLM process}
	\label{Fig:SelectionProcess}
\end{figure}

The search resulted in 341 publications from Scopus. We followed a selection process with 2 main stages in this SLM (Figure \ref{Fig:SelectionProcess}).

In stage 1, we applied the selection criteria (inclusion and exclusion criteria) over title, abstract and keywords, resulting in 62 papers (reduction of approximately 82\%). 1 paper was eliminated by EC1 (Study has no abstract); 10 by EC2 (Study without full text); 34 by EC3 (Study is not a primary study); 8 by EC4 (Study is not written in English) 3 by EC5 (Study is a copy or an older version of another study already considered); 2 by EC7 (Study with publication date prior to 2015); and 221 by EC8 (Study should contain problems related to smart contracts itself, problems about smart contract applications aren't enough). In stage 2, we applied the selection criteria considering the full text, resulting in 32 studies (a reduction of approximately 48\%). 1 paper was eliminated by EC2, EC3 and EC5; and 26 by EC8. From the first stage on, secondary works were also selected among the EC3 excluded papers. We separated those that didn't fit other exclusion criteria, obtaining 17 papers at the end of the 2nd stage. One more was included outside the search results because it was indicated by an expert due to its relevance, totaling 18 papers. The related work supported the determination of answers to our research questions through taxonomies, classifications and other contributions.

\section{Results and Discussions}\label{results-and-discussions}
The SLM study was carried out according to Section \ref{researchmethod}. Here we show the results for each of the research questions and discuss them. To answer the questions, an id was determined for each of the articles, as in~\cite{ourresultinpapers}.

To assist in the analysis and systematize the extracted data, categories for classifying the studies were defined, one for each research question. We considered the characteristics of the selected studies, both reusing categories already considered in the literature and defining new categories when necessary. It is possible that the same paper fits multiple categories. In cases of doubts about whether an article belongs to a certain category, we simply didn't include it. 

\subsection{Question 1: When and where have the studies been published?} 

\begin{figure}[!ht]
	\centering
		\includegraphics [width=1\textwidth]{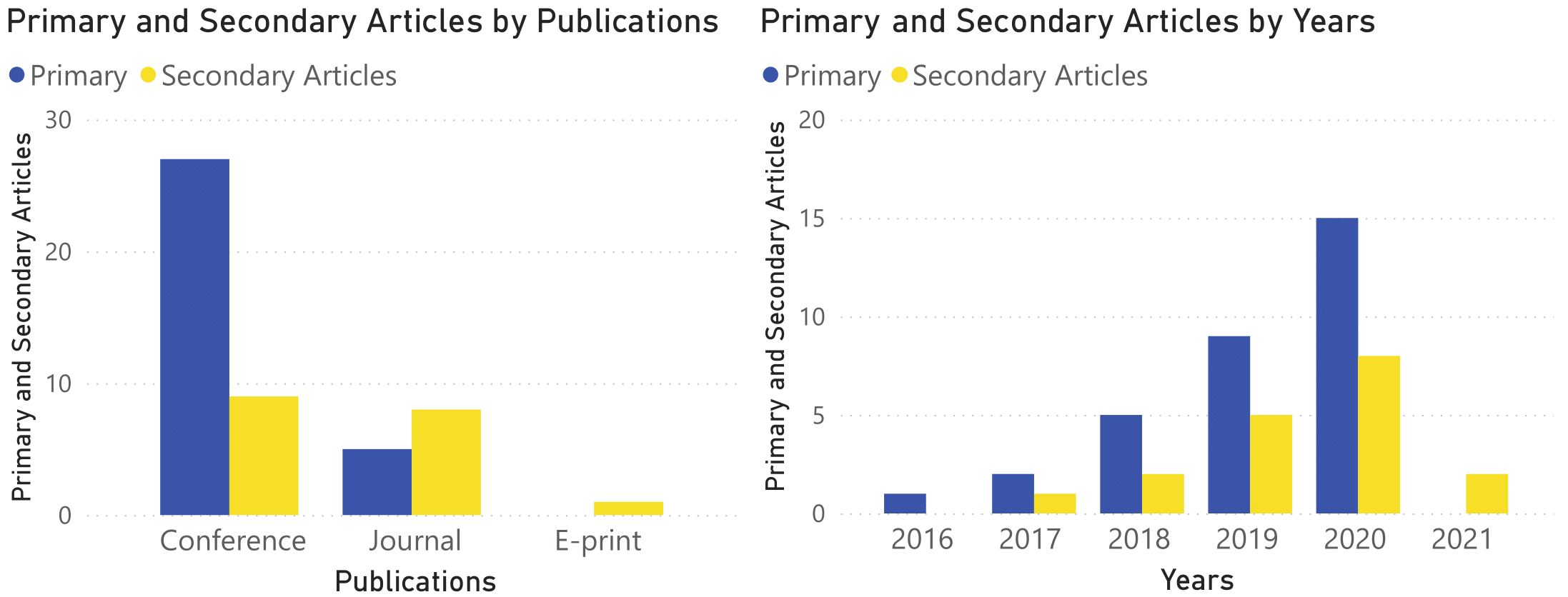}
		\caption{Frequency of year and source of publications of primary and secondary papers}
	\label{Fig:Year_Pub}
\end{figure}

Over the years, publications on smart contracts for Blockchain, had a growth similar to quadratic for both the primary and secondary papers, as presented in Figure \ref{Fig:Year_Pub}. We conducted our search in January 2021. So, we have only two papers this year, but considering that the period was less than a month, the number was not bad. Even with this growth rate, we find that the maturity of this field is still low. The majority, 72\% of all 50 papers were from conferences, and only 26\% from journals, is that in absolute terms it is still new compared to fields already established in the Information Technology. This is expected since it has not been even a decade since the first platform with support for SCs was inaugurated.

\begin{figure}[!ht]
	\centering
		\includegraphics [width=1\textwidth]{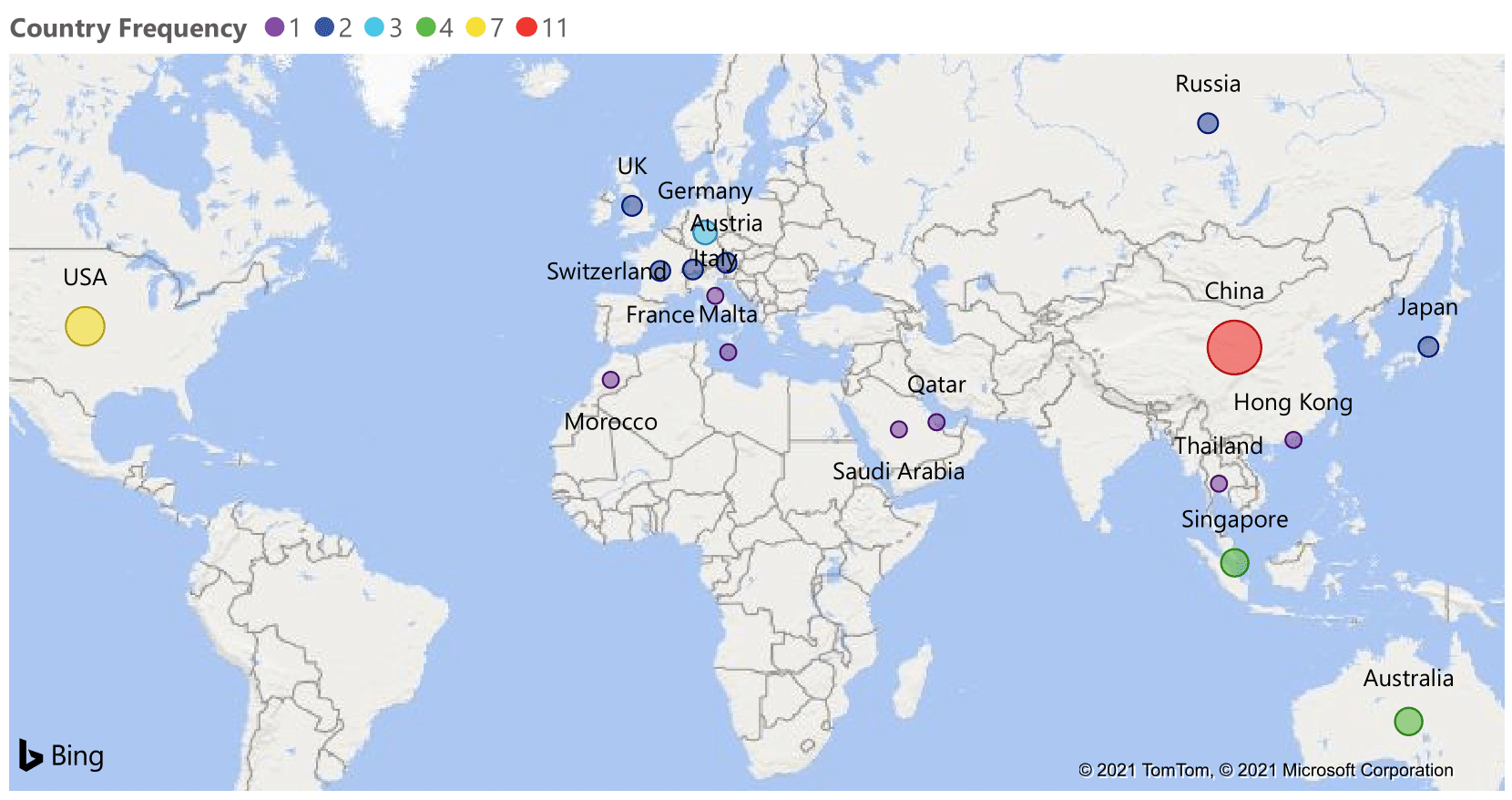}
		\caption{Countries of institutions where the authors of the primary papers are affiliated}
	\label{Fig:MapTest}
\end{figure}

We count the frequency of the countries from where the papers were written. They are shown in Figure \ref{Fig:MapTest}.
Countries that appeared only once (3.67\% each) were Italy, Malta,  Saudi Arabia, Hong Kong, Thailand, Qatar and Morocco; twice (6.67\% each) were Austria, Switzerland, United Kingdom,  France, Japan and Russia; three times (10\%) was Germany; four times (13.33\% each) were Singapore and Australia; seven times (23.33\%) was the USA; and eleven times (36.67\%) was China.

\subsection{Question 2: What are the problems/issues with respect to Smart Contracts?}

Most of the problems/issues pointed out in the literature were the exploitation of specific vulnerabilities in Ethereum. Solidity, the most used language in Ethereum, has several known problems\footnote[3]{The address \url{https://swcregistry.io/} has an updated list of known vulnerabilities.}.

In general, the articles seek to automatically find vulnerabilities by analyzing the source code, identifying the use of critical instructions (e.g., transfer funds), and analyzing the possibility of the code reaching these critical instructions.

The most common critical instructions mentioned were reentrancy, integer overflow, withdrawn and unprotected SELFDESTRUCT instruction. Besides security issues, some papers mention issues related to privacy [@8, @9, @11], performance [@5, @8, @9], governance and legality [@11].

Some articles studied problems not directly related to the source code of the SCs. They deal with external conditions of application execution, such as malicious or irrational use of applications, redesign of software engineering techniques to suit better the context of Blockchain, or monitoring of the use of decentralized applications (auditing).



\label{subsecrq1}

\subsection{Question 3: How are the problems/issues categorized or classified?}
\label{subsecrq2}

There was no common proposal to classify problems. Among primary papers, Luu\textit{ et al.} [\#1], for example, organized vulnerabilities based on their origin or cause, namely: dependence on the order of transactions, dependence on time, inadequate handling of exceptions and indentation. [\#15] used a similar classification. Another article by Petar Tsankov [\#4] organized the vulnerabilities as follows: insecure coding, unsafe transfers, unsafe inputs, transaction reordering, and reentrancy issues. Groce\textit{ et al.} [\#19] prefer to classify the vulnerabilities by their severity and the difficulty of being exploited.

Regarding secondary papers, Wang\textit{ et al.} [@17] provides an interesting taxonomy over SC security issues, also categorizing respective solutions for them, being the main problems: abnormal contract, program vulnerability and exploitable habitat. Huang\textit{ et al.} [@4] overview security themes from an SC lifecycle perspective. In each phase, vulnerabilities may be introduced or exploited, just as methods can be used to avoid them, being the phases: security design, security implementation, testing before deployment and monitoring and analysis. The systematization of 10 SC vulnerabilities classes based on Common Weakness Enumerator, provided by [@15], is worth mentioning.

Even though all the proposed problem classifications are different, they can be the basis to design a more complete taxonomy for the critical issues in SC. Also, there are examples where that issue and solution classifications fit well together [@17]. It might be worth considering combining them.

\subsection{Question 4: What are the proposed solutions to the problems/issues identified within the context of Smart Contracts?}\label{subsecrq3}

The material extracted for this question was organized considering related work that had categories or taxonomies for solutions\cite{tolmach2020survey}[@4, @13, @17]. Based on the resulting papers, we defined the categories as in Table \ref{Table:RQ4-distribution}, showing their distribution.

\begin{center}
\begin{longtable}[c]{l|c|c|c}
\caption{Distribution over the proposed solutions to the problems/issues identified}
\label{Table:RQ4-distribution}\\
\hline
\multicolumn{2}{c|}{\textbf{Methods}}
& \textbf{\#Primary Articles} & \textbf{Total (\%)} \\ \hline
\endfirsthead
\endhead
    \multirow{2}{*}{\textbf{Design}}
        & \multicolumn{1}{c|}{\textbf{Modeling}} & \multicolumn{1}{c|} {\makecell{1, 2, 3, 4, 5, 7, 8, 10,\\ 12, 14, 15,16, 20, 24,\\ 25, 28, 29, 31, 32}} & \multicolumn{1}{c}{59.38} \\\cline{2-4}
        & \multicolumn{1}{c|}{\textbf{Specification}} & \multicolumn{1}{c|} {\makecell{4, 7, 14, 15, 16,\\ 18, 24, 25}} & \multicolumn{1}{c}{25.00} \\\hline
    \textbf{Implementation} & \textbf{Design Pattern} & 6, 8, 11, 17 & 12.50 \\\hline

    \multirow{8}{*}{\textbf{Verification}}
        & \multicolumn{1}{c|}{\textbf{Theorem Proving}} & \multicolumn{1}{c|} {1, 5, 10, 18, 26, 27, 30} & \multicolumn{1}{c}{21.88} \\\cline{2-4}
        & \multicolumn{1}{c|}{\textbf{Model Checking}} & \multicolumn{1}{c|} {7, 15, 31} & \multicolumn{1}{c}{\textcolor{white}{0}9.38} \\\cline{2-4}
        & \multicolumn{1}{c|}{\textbf{Abstract Interpretation}} & \multicolumn{1}{c|} {2, 24} & \multicolumn{1}{c}{\textcolor{white}{0}6.25} \\\cline{2-4} & \multicolumn{1}{c|}{\textbf{Symbolic Execution}} & \multicolumn{1}{c|} {\makecell{1, 5, 10, 13, 14, 16,\\ 18, 19, 26, 30}} & \multicolumn{1}{c}{31.25} \\\cline{2-4} & \multicolumn{1}{c|}{\textbf{Runtime Verification}} & \multicolumn{1}{c|} {8} & \multicolumn{1}{c}{\textcolor{white}{0}3.13} \\\cline{2-4}
        & \multicolumn{1}{c|}{\textbf{Software Testing}} & \multicolumn{1}{c|} {6, 16, 19, 21, 32} & \multicolumn{1}{c}{15.62} \\\cline{2-4}
        & \multicolumn{1}{c|}{\textbf{Machine Learning}} & \multicolumn{1}{c|} {9, 16, 22, 23} & \multicolumn{1}{c}{12.50} \\\cline{2-4}
        & \multicolumn{1}{c|}{\textbf{Manual Auditing}} & \multicolumn{1}{c|} {19} & \multicolumn{1}{c}{\textcolor{white}{0}3.13} \\\hline

\end{longtable}
\end{center}

Regarding the proposed solutions, we only included papers that were certain of their classifications. The dominant category was Design Modeling with 19 papers, among them: Event-B, a set-based method [\#31]; State-Transition Systems like Colored Petri nets [\#28], Markov decision processes [\#16], Kripke structure [\#15], Dynamic Automata with Events [\#8] and state machines [\#7]; Abstract Syntax Tree-Level Analysis [\#12, \#25, \#29]; Control Flow Graph [\#1, \#2, \#4, \#5, \#10, \#12, \#14, \#16, \#20, \#24, \#32]; and Agent-based model [\#3], borrowed from Game Theory. Yet in Design realm, Specification has 8 papers,  among them: logic like Computation Tree Logic [\#7] and Linear Temporal Logic [\#15]; Horn Clauses [\#24]; Hoare-Style Properties [\#25]; and Program Path-Level Patterns like execution traces [\#14, \#16, \#18] and datalog [\#4].

In the Verification aspect, the most often approach is Symbolic Execution with 10 articles [\#1; \#5, \#10, \#13, \#14, \#16, \#18, \#19, \#26, \#30] followed by: Theorem Proving like Z3 [\#1, \#5, \#10, \#18], Isabelle/HOL [\#27] and Yices2 [\#30], with 7 articles in total; Software Testing, with approaches like unit testing [\#19], fuzzing [\#16, \#19], mutation testing [\#21] and Test Case Generation [\#32], with 5 papers; Machine Learning, among the approaches there are Random Forest [\#9], Imitation Learning [\#16], degree-free graph convolutional neural network [\#22], temporal message propagation network [\#22] and bidirectional long-short term memory with attention mechanism [\#23], totaling 4 papers; Model Checking, leveraging tools like NuSMV [\#7, \#15] and ProB [\#31], with 3 papers in total; Abstract Interpretation with 2 papers [\#2, \#24]; Runtime Verification [\#8] and Manual Auditing [\#19], each of them with 1 paper.

Note that it is common for Symbolic Execution approaches to leverage a satisfiability modulo theories solver, from the Theorem Prover category. Another consideration is that not only proposals for approaches were included in Table \ref{Table:RQ4-distribution}. We also include leveraged methods and tools used with experimental proposals. Hence, the Verification category contains 23 (71.875\%) papers,  Design has 20 (62.5\%) and Implementation 4 (12.5\%) papers [\#6, \#8, \#11, \#17], all belonging to its only subcategory, the Design Pattern.

\subsection{Question 5: Are there methods and tools available for the identified solutions?}\label{subsecrq4}

The material extracted for this question was schematized considering that the tool(s)/method(s): (i) are proposed; (ii) have implementation available (studies that no mention how to access or obtain the implementation, even if it exists, wasn't included); (iii) even if not proposed in the paper, are evaluated in experiments (reproducible cases will be mentioned); (iv) aren't proposed, but have relevant contributions; and (v) the possibility of conflict of interest is considerable (studies with less than half of the authors being part of a for-profit organization wasn't included). Table \ref{Table:RQ5-distribution} shows the distribution over this scheme.

\begin{longtable}[c]{c|c|c}
\caption{Distribution over methods and tools available for the identified solutions}
\label{Table:RQ5-distribution}\\
\hline
\textbf{ID} 
& \textbf{\#Primary Articles} & \textbf{Total (\%)} \\ \hline
\endfirsthead
\endhead
    \textbf{(i)} &
    \thead{1, 2, 3, 4, 5, 7, 8, 9, 10, 11, 12, 13, 14, 15,  16, \\ 17, 20, 21, 22, 23, 24, 25, 26, 27, 28, 30, 31, 32} &
    87.50 \\ \hline
    \textbf{(ii)} &
    1, 4, 5, 12, 16, 23, 24, 26, 30 &
    28.13 \\ \hline
    \textbf{(iii)} &
    \thead{1, 2, 5, 8, 9, 10, 12, 13, 14, 16, \\ 17, 18, 19, 21, 22, 23, 24, 28, 30, 32} &
    62.50 \\ \hline
    \textbf{(iv)} &
    6, 18, 19, 29 &
    12.50 \\ \hline
    \textbf{(v)} &
    7, 9, 12, 19, 29 &
     15.63 \\ \hline
\end{longtable}


As category (i) has a dominant frequency, with 29 papers (87.50\%), the category (iv) has the lowest representativeness, with only four papers (12.50\%). Even though the proposal for solutions is a category with a larger number, there are considerably fewer papers with available implementations, with only nine articles (28.13\%) in that category. About category (iii), it's worth mentioning papers that facilitate the reproduction of experiments/analyses [\#1, \#5, \#12, \#16, \#18, \#23, \#24, \#28]. We emphasize that we consider only those papers that have made available/indicated a dataset and tools/methods used to allow public access, being potentially reproducible.

Regarding the contributions presented in (iv) we highlight: the importance of Software Engineering in the area of Blockchain and SCs, being mentioned testing, design patterns and best development practices [\#6]. Regarding software testing, an important limitation pointed out in SC are the few options for execution environments for testing. In the case of the Ethereum platform there are only the main (the real), test (for developers) and simulation (local) networks [\#6]; In [\#18], we mention as the main contributions the security analysis framework to classify security issues in on-chain wallet contracts and the reproducible experiments; and, in [\#19], a very interesting flaws categorization, involving 22 categories, is used to outline 246 flaws found in audits. In addition, each of the flaws found is classified according to its severity (potential impact) and difficulty of exploitation. After an in-depth examination of empirical evidence, the authors came to interesting conclusions, such as: unit tests are ineffective in identifying failures, as the correlation between the number of pre-existing unit tests and the audit results was weak; there's a trade-off between cost/degree of automation and flaws detection with the exploitation of high severity and low difficulty; and many failures can be solved by adopting the ERC20 standard.

In relation to category (v), it is clear that identifying cases of conflict of interest having as criterion that at least half of the authors of an article are part of a for-profit organization is not really a good decisive criterion. Identifying cases like these is not a simple task, after all, conflicting interests are not necessarily financial, even though the existence of a conflict of interest in some cases may be evident. On the other hand, how much this affects the content is generally not clear. Then, category (v) indicates only the possibility of conflict of interest and might be worth remembering this detail during reading. 

\subsection{Discussion}\label{discuss}

Summarizing what was seen in the studies addressed, we can understand that the observed vulnerabilities have as common cause the difficulty of detecting incompatibilities between the intended and the real behavior. This is largely due to the platform's immaturity, which still has high-level resources with complicated and unexpected low-level behaviors for conventional developers. But the technology is considerably new, and it is expected to mature.

Beyond this scenario, in which there is an increase in attacks exploiting vulnerabilities in SCs, and the existence of already published vulnerable contracts is known, Almakhour \emph{et al.} [@13] mention three reasons to apply formal specification and verification to SCs before their deployment: once an SC is published, any vulnerabilities in its code will become permanent; programmers' lack of knowledge about proper programming semantics to reflect high-level workflow may lead to misconception issues, resulting in ``unfair contracts"; and, many programming paradigms used to develop SCs were not designed to be used in the context of the Blockchain environment~\cite{kalra2018zeus}[@13]. Besides, Liu and Liu [@5] recommended carrying out more research to design more complete formal verification tools, combining formal verification methods with vulnerability analysis methods that complement each other. Observing the results presented in Table \ref{Table:RQ4-distribution}, Modeling and Specification have been extensively studied. The solutions least explored in the research, that were related to Formal Verification are: Runtime Verification (3.13\%), Abstract Interpretation (6.25\%), and Model Checking (9.38\%). It's an indicative that little was exploited of such techniques to deal with vulnerabilities in SCs, which points us to a direction to be investigated.

A limitation often mentioned in relation to some methods, e.g. symbolic execution and model checking, is path/state explosion [@13, @18], a situation predominantly caused by unbounded iterations. It's important to remember that, in practice, there are no unbounded iterations because there is a well-defined limit: the gas. Therefore, an adequate modeling of the gas usage by SCs is a promising way to mitigate this limitation of the methods. In this case, simple heuristics can be applied to define values for variables such as the balance of SCs and the cost per instruction.

Regarding to software testing approaches, a considerable part of them have a dynamic nature, requiring the execution of system under test. We known that there is no suitable environment for dynamic analysis. Those that are available range from simulated or real network environments, but each of them with own drawbacks. The resulting papers that mention software testing often wasn't clear if the context was static or dynamic, so we suggest that future work in this topic should address these details.

With respect to reasoning on security vulnerabilities, attacks often are based on exploitation of more than one vulnerability. In this sense there is a certain similarity between the attacks and the propagation [@15]. In addition, it is highly likely that new vulnerabilities will be discovered only after they have already been exploited, and only after that, low-level properties may be defined to mitigation. So, one may consider anticipating new attacks to handle them. One possibility may be to consider the issue of finding new attacks as an optimization problem where the search space are the infinite  combinations of vulnerabilities, applying some metrics, like gas cost.

\section{Threats to Validity}\label{threats}
We clarify some of the possible limitations of this mapping. During the process of selecting studies and extracting data, there was the matter of subjectivity. One of the solutions to avoid subjectivity was that all the authors participated in defining the search string. When the papers were selected to be fully read, the authors conducted a check by reading full papers of the other three authors.



The use of only one electronic database causes a limitation in the set of relevant papers that can be obtained as a result. However, as mentioned previously, Scopus has been selected as it also contains most of the publications from other databases. One other limitation is the lack of snowballing (forward and backward)~\cite{jalali2012systematic} which could have brought some more relevant publications to complement and to bring a more qualified analysis. Looking directly at the publications of some research groups also is a valid search to enhance our process. These two aspects might have limited the input of more papers. Also, regarding the tertiary analysis provided, a limitation is the fact that we did not use specific strings to find secondary papers, so some relevant articles may have been left out. However, this is the initial study being performed and definitely these aspects will be considered in the continuity of this mapping.

\section{Conclusions}\label{conclusion}

A mapping study gives an idea, in the early phases, of shortcomings in existing evidence, which becomes a basis for future investigations \cite{kitchenham2011using}. This paper
presented a systematic mapping on vulnerabilities and open issues of Smart Contracts in the Blockchain context. We have analyzed 32 primary references and 18 secondary articles. The findings from this study are expected to contribute to the existing knowledge about the vulnerabilities of SCs. In summary, we concluded that: (i) critical instructions that lead to problems directly related to the correct writing of contracts are:  reentrancy, integer overflow, withdrawn and unprotected SELFDESTRUCT instruction; (ii) It was not possible to identify a standard proposal for classification of problems, both in primary and secondary works. Even though the classification is different, it is possible to use them as a basis to design a complete taxonomy regarding the critical issues in SCs; (iii) the papers showed that the proposed solutions to the identified problems are: Design, Implementation, and Verification, the most used methods, Modeling (present in 59.38\% of the papers), which belongs to the Design category and Symbolic Execution (present in 31.25\% of the papers), which belongs to Verification. Also, some studies point to the need for more research to design formal verification tools, combining formal verification methods with vulnerability analysis.

Thus, as future work, we intend to update this SLM to include more electronic databases and include more stages in the selection process, such as snowballing and research groups. Furthermore, extending the scope of the mapping to more categories related to SCs security problems/issues, as in the categorizations proposed by works like [@16, @17] and traditional software vulnerability analysis ~\cite{ji2018coming, shen2020survey} is a promising path. As future work, we can also suggest: research in the perspective of machine learning on the application of automatic vulnerability detection for SCs, since it is a promising vulnerability detection technique even for traditional software~\cite{ji2018coming,shen2020survey}, does not rely heavily on rules defined by human experts and has obtained competitive results; research proposing solutions involving dynamic analysis or software testing, seeking to deal with the limitations of the execution environment. 


\section*{Conflicts of Interest}
The authors declare that they have no conflicts of interest.

%
%
%

\bibliographystyle{splncs04}
\bibliography{general_bibliography,selected_studies_bibliography}

\end{document}